\documentclass[pre,twocolumn,showpacs]{revtex4}
\usepackage{graphics,graphicx,dcolumn,bm,fleqn,epic,eepic,float}
\usepackage{amssymb,amsmath,multirow,rotate,color}
\usepackage{subfigure}

\def \bsigma {\mbox {\boldmath $\sigma$}}
\def \bxi {\mbox {\boldmath $\xi$}}
\def \bm {\mbox {\boldmath $m$}}
\def \bA {\mbox {\boldmath $A$}}
\def \bS {\mbox {\boldmath $S$}}

\begin{document}

\setcounter{page}{0}
\input epsf

\title{\bf Pattern reconstruction and sequence processing
in feed-forward layered neural networks near saturation}
\author{F. L. \surname{Metz} and W. K. \surname{Theumann}}
\affiliation{Instituto de F\'\i sica, Universidade Federal do Rio
Grande do Sul, Caixa Postal 15051, 91501-970 Porto Alegre, Brazil}

\date{\today}
\thispagestyle{empty}

\begin{abstract}

The dynamics and the stationary states for the competition between
pattern reconstruction and asymmetric sequence processing are
studied here in an exactly solvable feed-forward layered neural
network model of binary units and patterns near saturation.
Earlier work by Coolen and Sherrington on a parallel dynamics far
from saturation is extended here to account for finite stochastic
noise due to a Hebbian and a sequential learning rule. Phase
diagrams are obtained with stationary states and quasi-periodic
non-stationary solutions. The relevant dependence of these
diagrams and of the quasi-periodic solutions on the stochastic
noise and on initial inputs for the overlaps is explicitly
discussed.

\end{abstract}


\maketitle
\setcounter{page}{1}

\section{Introduction}

Models of attractor neural networks for processing sequences of
patterns, as a realization of a temporal association, have been of
great interest following Hopfield's pioneering work [1-7] and
renewed interest has come through both, the availability of new
theoretical dynamic approaches to study the evolution of
disordered systems, in particular neural networks near saturation
[8-12] and experimental findings. Among the latter are the results
of Miyashita {\it {et al.}} who showed that serial positions of
stimuli to which monkeys were exposed during training are
converted to spatial correlations of neural activities
\cite{MC88,Mi88}.

A neural network model with a symmetric learning rule that
consists of a Hebbian part and a pair of pattern sequences has
been proposed to interpret the experiments \cite{GTA93,Br94}. The
analysis of the equilibrium states for both finite and extensive
loading of patterns, revealed the presence of correlated
attractors with a decreasing correlation as the separation of the
patterns in the sequence increases, in apparent support of the
experimental work. Phase diagrams were obtained where correlated
attractors compete with Hopfield-like attractors and with
symmetric states \cite{Cu93,CT94}. The dynamical evolution to the
stationary states has been studied recently by means of dynamical
replica theory \cite{UHO04}.

Neural network models for sequence processing with asymmetric
interactions are more natural from a biological point of view. Due
to the lack of microscopic detailed balance, however, the
equilibrium states of the network cannot be obtained by means of
the statistical mechanics approach \cite{AGS87}, and one has to
resort to a dynamical study. Analytic and numerical studies of the
stationary states and some aspects of the dynamics for the
competition between pattern reconstruction and asymmetric sequence
processing in the case of finite loading of patterns, that is in
the absence of stochastic noise, where the ratio $\alpha=p/N$
between the number of patterns $p$ and units $N$ is zero, appeared
in works by Coolen {\it{et al.}} \cite{CS92,WSC95}. Phase diagrams
for the stationary states of a parallel dynamics exhibit either
stable Hopfield-like or symmetric mixture states characterized by
fixed-point solutions and stable limit cycles described by
periodic fixed points. Non-stationary solutions of the dynamics
also appear for increasingly large number of patterns \cite{CS92}.
More recent works deal with stationary limit cycles in asymmetric
sequence processing without pattern reconstruction, for extensive
loading of patterns [12,23-26].

It is important to consider the effects of extensive loading of
patterns and to study the stability of the states that appear in
the competition between pattern reconstruction and asymmetric
sequence processing to the presence of stochastic noise (finite
$\alpha$). This noise could be due to the presence of a previously
learnt macroscopic number of patterns following a specific
learning rule. Since the problem is a dynamical one, it is also
important to study the dependence of the network performance on
initial conditions.

Those are issues that, apparently, have not been considered before
and the purpose of the present paper is to study them on a
tractable dynamical feed-forward layered neural network model with
no feed-back loops \cite{DKM89}. This is an extensively used model
that consists of identical layers of $N$ non-interacting units on
each layer, with synaptic interactions only between units on
consecutive layers. The feed-forward nature of the model and the
updating of the units on each layer endow the network of a
dynamics in which the layer index becomes a discrete time. We
consider an interaction matrix that yields either a static Hebbian
noise or a dynamic Hebbian plus-sequential noise and restrict the
work to binary units and patterns. In the finite loading case, the
model becomes identical to that of Coolen and Sherrington for a
parallel dynamics \cite{CS92}.

The outline of the paper is the following. In Sec. 2 we present
the model and the relevant order parameters. In Sec. 3 we derive
the macroscopic dynamics of the network and in Sec. 4 we present
the results in the form of phase diagrams of stationary states and
regions of non-stationary solutions. We end with a discussion and
conclusions in Sec. 5.

\section{The model}

The network model consists of $L$ layers with $N$ binary Ising
units (neurons) on each layer $l$ in a microscopic state
$\bsigma(l)=\{\sigma_1(l),\dots,\sigma_N(l)\}$, in which each
$\sigma_i(l)=\pm 1$. The state $+1$ represents a firing neuron and
the state $-1$ a neuron at rest. The microscopic dynamics of the
network is generated as follows: given a configuration on the
first layer, $\bsigma(1)$, all units on layer $l+1$ are updated
simultaneously according to the alignment of each unit $i$ to its
local field
\begin{equation}
h_i(l+1)=\sum^{N}_{j=1}J_{ij}(l)\sigma_j(l) \,\,\,\,,
\label{1}
\end{equation}
due to the states of the units on the previous layer $l$,
following a stochastic law with probability
\begin{equation}
\rm Prob(\sigma_i(l+1)|\bsigma(l))
=\frac{\exp[\beta\sigma_i(l+1)h_i(l+1)]}{2 \cosh[\beta
h_i(l+1))]}\,\,\,\,.
\label{2}
\end{equation}
Thus, the network has a parallel dynamics with no feed-back loops
in which the layer indices may be associated with discrete time
steps. Here, $J_{ij}(l)$ is the synaptic connection defined below
between unit $j$ on layer $l$ and unit $i$ on layer $l+1$. The
parameter $\beta=T^{-1}$ controls the synaptic noise such that the
dynamics of the network becomes a deterministic one when
$T\rightarrow 0$ and fully random when $T \rightarrow \infty$. In
the former case, the state of a unit is given by the deterministic
form
\begin{equation}
\sigma_i(l+1)={\rm sgn}[h_i(l+1)] \,\,\,.
\label{3}
\end{equation}

A macroscopic set of $p=\alpha N$ independent and identically
distributed random patterns $\bxi^{\mu}(l)
=\{\xi_1^{\mu}(l),\dots,\xi_N^{\mu}(l)\}$, $\mu=1,\dots,p$, and
each $\xi_i^{\mu}(l)=\pm 1$ with probability $\frac{1}{2}$ are
stored in the learning stage on the units of each layer $l$,
independently of other layers. Since we are interested in the
retrieval of one or a small number $c$ of patterns and in the
recognition of a finite sequence, we assume a learning rule of the
form
\begin{eqnarray}
J_{ij}(l)&=&\frac{1}{N}\sum_{\mu,\rho=1}^c\xi_i^{\mu}(l+1)A_{\mu
\rho} \xi_j^{\rho}(l)\\ \nonumber
&+&\frac{1}{N}\sum_{\mu,\rho=c+1}^p\xi_i^{\mu}(l+1)B_{\mu\rho}
\xi_j^{\rho}(l) \,\,\,\,,
\label{4}
\end{eqnarray}
where
\begin{eqnarray}
A_{\mu \rho}&=&\nu \delta_{\mu \rho}+(1-\nu)
S_{\mu \rho}\,\,,\,\,\,\,\,\mu\,\, {\rm{mod\,\, c}}\,\,\,,\\
\nonumber B_{\mu \rho}&=& b \delta_{\mu \rho}+(1-b)S_{\mu \rho}
\,\,,\,\,\,\,\,\mu\,\, {\rm{mod\,\, p}}\,\,\,, \label{5}
\end{eqnarray}
in which $B_{\mu \rho}$ is non-zero only for $\mu\geq c+1$. These
are the (layer independent) elements of matrices $\mathbf{A}$ and
$\mathbf{B}$, with continuous independent $\nu$ and $b$
($0\leq\nu,b\leq 1$), in which $\delta_{\mu \rho}=1$ if
$\mu=\rho$, and zero otherwise, while $S_{\mu
\rho}=\delta_{\mu,\rho +1}$ is a permutation matrix element. The
signal-to-noise analysis carried out below depends on the
assumption that the patterns can be separated into $c$ "low" or
condensed ones, with finite overlaps with the states of the
network in the $N\rightarrow\infty$ limit and ($p-c$) "high" or
non-condensed patterns with overlaps of $O(1/\sqrt N)$ with the
states. The assumption is justified in the case of a block-matrix
interaction (here, a diagonal two-block matrix) and the specific
form we choose is inspired in an earlier proposal \cite{CT94}. In
that spirit we assume here not necessarily the same weights, $\nu$
and $b$. This is in order to contemplate different simple
possibilities and to avoid an eventual closure problem in the
noise term discussed below. The first part of $J_{ij}(l)$
represents a finite cycle which is a superposition of a Hebbian
learning (the diagonal part of $\mathbf{A}$) of single patterns
and of patterns in a sequence (the non-diagonal part). Again, in
the spirit of that reference, one may think of the condensed part
in the presence of an infinite cycle (in the large $p$ limit) of a
previously learnt similar superposition of Hebbian and sequential
patterns represented by the second part of $J_{ij}(l)$, which acts
as a noise. Eventually, one could have $0\leq\nu\leq 1$ and a
Hebbian noise with $b=1$ \cite{CT94}, which is a rather convenient
choice.

The learning stage of the network is a dynamical process that
involves patterns on two consecutive layers. The Hebbian part of
the rule may be thought as a static process that reinforces the
learning of the same pattern on every pair of consecutive layers
(times) whereas the sequential part of the rule is a dynamic
process in which the synaptic interaction is due to a pattern at a
given time with the following pattern in the sequence at the next
discrete time.  It has been found before, in studying the finite
loading case, that the competing static process has a stabilizing
effect on the dynamic sequential process leading to a phase of
symmetric states, which are the only ordered stable states within
a wide range of relative synaptic strengths under an appropriate
amount of synaptic noise parameter $T$. Eventually, the static
process may fail to lock the transitions in the dynamic process
and non-stationary quasi-periodic solutions may appear which have
already been found in the case of finite loading \cite{CS92}.
These are natural features of the model that are enhanced by
stochastic noise (finite $\alpha$) as will be seen and discussed
in this work.

The properties of $\mathbf{A}$, in particular its eigenvalues and
the complete set of orthogonal eigenvectors, lead to symmetries of
the solutions of the non-linear parallel dynamics in the form of a
$\nu/(1-\nu)$ duality, that has been discussed for finite loading
of patterns \cite{CS92}. We find here an additional $b/(1-b)$
duality. For the case of continuous bifurcations of solutions, the
eigenvalues of the matrix yield the transition temperatures to the
ordered states and the eigenvectors give the symmetry directions
of the macroscopic overlaps between the states of the network and
the patterns.

We define the macroscopic overlap between the configuration
$\bsigma(l)$ of the network on layer $l$ and one or more condensed
key patterns $\bxi^{\mu}(l)$,\, $\mu=1,\dots,c$, on that layer as
the large-N limit, $m^{\mu}(l)$, of
\begin{equation}
m^{\mu}_{N}(l)=\frac{1}{N}\sum_{i=1}^{N}\xi_i^{\mu}(l)
\langle\sigma_i(l)\rangle\,\,\,,
\label{6}
\end{equation}
where the brackets denote a thermal average with Eq.(2). Since the
number of condensed patterns is finite, one may use the
self-averaging property to write $m^{\mu}(l)=\langle\xi_i^{\mu}(l)
\langle\sigma_i(l)\rangle\rangle$, where the outer bracket here
and below denotes a configurational average over the patterns.
Similarly, we define the overlap between the same configuration
and a given non-condensed pattern $\bxi^{\mu}(l)$,
$\mu=c+1,\dots,p$, on layer $l$,
\begin{equation}
M^{\mu}_{N}(l)=\frac{1}{N}\sum_{i=1}^{N}\xi_i^{\mu}(l)
\sigma_i(l)\,\,\,.
\label{7}
\end{equation}
Assuming that a given configuration of the first layer has a
finite overlap $m^{\mu}(1)=O(1)$ with one or more condensed
patterns and overlaps $M^{\mu}_{N}(1)=O(1/\sqrt{N})$ with the
non-condensed patterns, the dynamic evolution of the network will
yield overlaps $m^{\mu}(l)=O(1)$ and
$M^{\mu}_{N}(l)=O(1/\sqrt{N})$ on the following layers. We
consider next the evolution equations for the overlaps.

\section{Dynamics of the network}

Adapting the standard procedure for the layered network to our
model, we write the local field at a unit on layer $l+1$ due to
the overlaps with all patterns on layer $l$, in the large-N limit
\cite{DKM89},
\begin{equation}
h_i(l+1)=\sum_{\mu,\rho=1}^c\xi_i^{\mu}(l+1)A_{\mu \rho}
m^{\rho}(l)+ z_i(l)\,\,\,,
\label{8}
\end{equation}
where the first term is the signal and $z_i(l)$ is the large-N
limit of the noise
\begin{equation}
R_i(l)=\sum_{\mu,\rho=c+1}^{p}\xi_i^{\mu}(l+1)B_{\mu
\rho}M^{\rho}_{N}(l)\,\,\,,
\label{9}
\end{equation}
due to the overlaps of the states with the non-condensed patterns.
This is a random quantity in both the patterns on layer $l+1$ and
the implicit dependence on thermal and configurational randomness
of the overlaps $M^{\mu}_{N}(l)$ on layer $l$ due to the previous
layers.

The noise gives a finite contribution to the local field. Indeed,
due to the fact that $M^{\mu}_{N}(l)=O(1/\sqrt{N})$ and that
$R_i(l)$ is a sum of a large number of statistically independent
random variables, one can first apply the central limit theorem
and then the law of large numbers to conclude that $z_i(l)$
follows a Gaussian distribution with mean zero and a variance
$\Delta^{2}(l)$ given by the large-N limit of
\begin{equation}
\Delta^{2}_N(l)=\sum_{\mu=c+1}^{p}\langle\langle[bM^{\mu}_{N}(l)
+(1-b)M^{\mu-1}_{N}(l)]^{2}\rangle\rangle
 \,\,\,.
\label{10}
\end{equation}
We used the fact that the patterns are unbiased and uncorrelated
random variables with $\langle\xi_i^{\mu}(l+1)\rangle=0$ and
$\langle\xi_i^{\mu}(l+1)\xi_i^{\nu}(l+1)\rangle =\delta_{\mu\nu}$.

We make use now of the local field to derive the recursion
relations for the vector overlaps with the condensed patterns,
$\bm(l)=\{m^{\mu}(l)\}, \mu=1,\dots,c$, and for the variance of
the noise. For the former we obtain
\begin{equation}
\bm(l+1)= \langle\bxi(l+1)\int
Dz\,\tanh\{\beta[\bxi(l+1).\bA\bm(l)+\Delta(l)z]\}\rangle\,\,\,,
\label{11}
\end{equation}
where $Dz=e^{-z^{2}/2}dz/\sqrt{2\pi}$ and the brackets denote an
average over the explicit patterns. The variance of the noise
requires recursion relations not only for the average squared
non-condensed overlaps, $\langle\langle
M^{\mu}_{N}(l)^{2}\rangle\rangle$ and $\langle\langle
M^{\mu-1}_{N}(l)^{2}\rangle\rangle$, which can be derived in the
usual way \cite{DKM89}, but also for the correlation of two
consecutive overlaps $\langle\langle M^{\mu}_{N}(l)
M^{\mu-1}_{N}(l)\rangle\rangle$. This generates, in turn,
correlations between next-to-consecutive overlaps, $\langle\langle
M^{\mu}_{N}(l) M^{\mu-2}_{N}(l)\rangle\rangle$ and so on, which
requires to keep track of a general form
\begin{eqnarray}
C_{n}^2(l)&=&\sum_{\mu=c+1}^{p}\langle\langle[bM^{\mu}_{N}(l)
+(1-b)M^{\mu-1}_{N}(l)]\\ \nonumber &.&[bM^{\mu-n}_{N}(l)
+(1-b)M^{\mu-n-1}_{N}(l)]\rangle\rangle
 \,\,\,
\label{12}
\end{eqnarray}
leading altogether to $p-c+1$ recursion relations,
\begin{eqnarray}
\Delta^{2}(l+1)&=& {\tilde b}^2(\alpha+{\beta}^2I^2\Delta^2(l))\\
&+& 2b(1-b){\beta}^2I^2C_{1}^2(l)\,\,\,,\nonumber\\
C_{1}^2(l+1)&=& {\tilde b}^2{\beta}^2I^2C_{1}^2(l) \\
&+& b(1-b)[\alpha+{\beta}^2I^2(\Delta^2(l) +C_{2}^2(l))]\,\,\,,\nonumber\\
C_{n}^2(l+1)&=& {\tilde b}^2{\beta}^2I^2C_{n}^2(l)\\
&+& b(1-b){\beta}^2I^2(C_{n-1}^2(l) +C_{n+1}^2(l))\,\,\,,\nonumber
\label{13}
\end{eqnarray}
for $n=2,\dots,p-c-1$ and $C_{p-c}=\Delta$, where
\begin{eqnarray}
{\tilde b}^2&=&b^2+(1-b)^2 \\ \nonumber I(l) &=&1-q(l)\,\,\,.
\label{14}
\end{eqnarray}
Here, $q(l)=\langle\langle \sigma(l)\rangle^{2}\rangle$ is the
spin-glass order parameter given by
\begin{equation}
q(l)=\langle\int Dz\,
\tanh^{2}\{\beta[\bxi(l+1).\bA\bm(l)+\Delta(l) z]\}\rangle\,\,\,.
\label{15}
\end{equation}

Note that the Gaussian noise is symmetric under the change $b$
$\rightarrow (1-b)$ and that in the case where either $b$ is one
or zero, that is for purely Hebbian or sequential noise,
respectively, the recursion relation for the variance reduces to
the simple form
\begin{equation}
\Delta^{2}(l+1)=\alpha + \beta^{2}I^{2}(l)\Delta^{2}(l)\,\,\,.
\label{18}
\end{equation}
Otherwise, one has to face the full set of recursion relations
which may become quite a task for an asymptotically large $p$.
Indeed, the system of relations may not form a closed and finite
set and there is no guaranty that this is not, in general, the
case. Fortunately, by working out numerically the equations for
the dynamics in all the cases we were interested in this work we
found that the set of recursion relations is practically finite
and one proceeds as follows.

Given an initial overlap $\bm(1)=\{m^{\mu}(1)\},\, \mu=1,\dots,c$,
on the first layer, and taking $\Delta^{2}(1)=C_{n}^2(1)=\alpha$,
for all $n$, Eqs. (11)-(17) describe the dynamics of the network
and its stationary states. The latter are given by the fixed-point
solutions $\bm^*=\bm(l+1)=\bm(l)$,
$\Delta^*=\Delta(l+1)=\Delta(l)$ and
$C_{n}^*=C_{n}(l+1)=C_{n}(l)$, for each $n=1,\dots,p-c-1$.
Expressing $C_{1}^*$ in terms of all higher $C_{n}^*$'s yields
$(\Delta^*)^2\propto \alpha$ and in the case of $b$ either one or
zero this reduces to the simple form
\begin{equation}
(\Delta^*)^{2}=\frac{\alpha}{1-\beta^{2}(1-q)^{2}}\,\,\,,
\label{19}
\end{equation}
where $q$ is the fixed-point value of $q(l)$.

There may also exist other solutions, as non-stationary states, as
will be seen in the next section. Solving for $T=0$, searching for
fixed-point solutions and for non-stationary states one finds that
only a small number of the $C_{n}^*$'s on the $(p-c)$-cycle of
non-condensed overlaps are clearly non-zero and all others vanish.
The ones that survive become smaller with increasing $n$
indicating the vanishing of the averaged overlaps $\langle\langle
M^{\mu}_{N}(l) M^{\mu-n}_{N}(l)\rangle\rangle$.

In the case of finite loading of patterns, where $\alpha=0$, the
variance of the noise vanishes and the equations for the overlaps
and the spin-glass order parameter become disconnected and we
recover the equations for the overlaps found in previous work
\cite{CS92}.

\section{Results}

We show and discuss next our results and use the solutions for
finite loading as a guide. We are mainly interested in the network
performance in the case of training with a macroscopic number of
patterns and on the impact of stochastic noise on the dynamics and
the stationary states. Among the important features are the
critical loading level, the relative size of the regions with
meaningful information processing, the presence of stable
symmetric mixture states and quasi-periodic non-stationary
solutions. We also want to find out the specific dependence, if
any, of these properties on the form of the stochastic noise and
we restrict ourselves to Hebbian or Hebbian plus sequential noise.

The phase diagrams discussed below are obtained from the dynamical
equations for the macroscopic order parameters and they depend
naturally on initial conditions. Quite different sets of these
conditions may reflect distinct basins of attraction of the
stationary states of the model (specified by a given pair of
values of $\nu$ and $b$) for given $T$ and $\alpha$. It will be
seen that non-stationary states can be reached from certain
initial overlaps.

We concentrate on a small number of macroscopic condensed
overlaps, specifically $c=3$ and $c=4$, since the results depend
on whether $c$ is even or odd, with the remaining $p-c$ patterns
as noise in the case of extensive loading. This already
illustrates the typical results. For specific results on higher
values of $c$ and the differences between an even and odd number
of patterns in the case of finite loading we refer the reader to
earlier work \cite{CS92}. Some of these results will be used below
to infer the changes that one may expect for extensive loading.

\subsection{Finite loading}

First we reconsider the solutions for finite loading, where
$\alpha=0$, in order to show the phases that appear, the periodic
features of the cyclic phase and the quasi-periodic solutions in
the region of unstable fixed points.

The $(T,\nu)$ phase diagram of stationary solutions for $\alpha=0$
and $c=4$ is shown in Fig.1 for the best initial overlaps that
favor single-pattern reconstruction (Hopfield or Hopfield-like
solutions, see below). There is a paramagnetic phase (P) where
$\bm^*=0$ and $q=0$ above $T=1$. This is a line of continuous
bifurcations to a phase of stable symmetric fixed-point solutions
(S), $m_1^*=m_2^*=m_3^*=m_4^*\neq 0$, that ends to the left at a
boundary (dashed line) obtained analytically beyond which the
symmetric states become unstable. This is a region of
non-stationary quasi-periodic solutions marked q-p, as will be
seen below, that is included with the warning that it is not a
true part of the phase diagram of stationary states. The same
applies to all other phase diagrams in this paper where the q-p
solutions are shown. It is worth noting that the symmetric states
are the only stable states in the $S$ region due to the
stabilizing effect of the static condensed patterns in the
learning rule and that this is a characteristic feature of the
model, as discussed in Sec. 2. Note that the symmetric states are
enhanced by synaptic noise.

Eventual alterations of the learning rule aimed at eliminating
what has been termed as "spurious" (here not at all) symmetric
mixture states, say by means of the introduction of a bias in the
distribution of the patterns \cite{AGS97/2}, would leave nothing
but the non-stationary q-p states over a large region of the phase
diagram.

The upper region of the phase diagram is a phase of stable
Hopfield-like fixed-point solutions (H), with the four components
$m_1^*,\dots,m_4^*$ not all equal and different from zero, except
at $T=0$ where one component may be one and the other ones zero.
Using this zero $T$ result as an initial solution we construct by
numerical iteration of the overlaps the upper and lower phase
boundaries indicated by solid lines. All other initial conditions
yield phase boundaries above the upper or below the lower line and
larger regions of q-p solutions. Those lines are boundaries of
first-order transitions for all values of $\nu$, except $\nu=1$ or
zero. The first-order transition is to the region of q-p solutions
with remanent finite overlap components. The location of the
transition for $c=4$ is, as expected, close to that for higher $c$
obtained in ref. \cite{CS92}. We also find, as expected, larger
discontinuities at the first-order transition and shorter remanent
overlaps in the q-p region, with our smaller number of condensed
patterns than in that reference.

The lower region of the phase diagram is the phase of period-4
stable stationary cyclic solutions (C). These are solutions in
which $m_i(l+4)=m_i(l)\neq 0$ for $i=1,\dots,4$, with one overlap
close to one and the other ones near zero in each time step after
a transient. This fact characterizes sequence processing in which
the network makes a transition from one pattern to the next at
each layer. The upper and lower first-order phase boundaries
appear symmetrically in the phase diagram due to the $\nu/(1-\nu)$
duality and both are equally sensitive to initial values for the
overlaps.

\begin{figure}
\centering
\includegraphics[width=7cm,height=6cm]{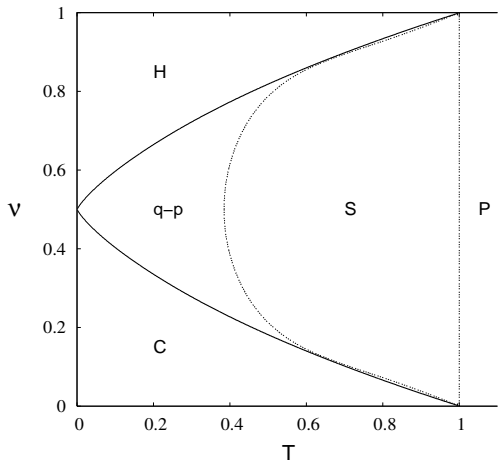}
\caption{Phase diagram for $c=4$ condensed patterns for finite
loading ($\alpha=0$), with initial overlaps
$m_{1}=1,m_{2}=m_{3}=m_{4}=0$. Hopfield-like (H), symmetric (S),
cyclic (C), paramagnetic (P) phases and region of quasi-periodic
(q-p) solutions. Solid and dashed lines indicate first-order and
continuous transitions, respectively, here and below.}
\label{fig1}
\end{figure}

The nature of the cyclic and quasi-periodic solutions for
$\alpha=0$ and $c=4$ is best illustrated by the power spectra
shown in Figs. 2 (a) and (b), respectively. The first one is for a
typical $\nu=0.1$ and $T=0.15$, within the phase of cyclic states,
and the other one is for a typical $\nu=0.30$ and $T=0.35$ in the
region of non-stationary states. The power spectrum $S(\omega)$,
where $\omega$ is the frequency conjugate to the layer index (a
discrete time) may be defined in extension to the case of a
continuous time \cite{ER85}, as
\begin{equation}
S(\omega)=({\rm
{const})}\,{\rm{lim}}_{L\rightarrow\infty}\frac{1}{L}
|\sum_{l=0}^{L}e^{i{\omega}l}m_{1}(l)|^{2}\,\,\,,
\label{18}
\end{equation}
where $m_{1}(l)$ is any one of the components of the overlaps.

\begin{figure}
\centering
\includegraphics[width=7cm,height=12cm]{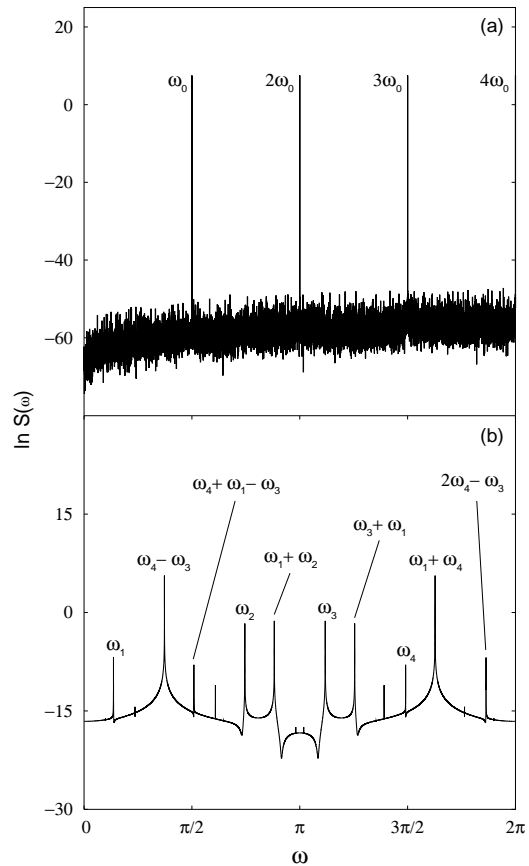}
\caption{Power spectrum $S(\omega)$, for $c=4$, and $\alpha=0$.
(a) Periodic 4-cycle for $\nu=0.1$ and $T=0.15$ in the C phase,
with $\omega_{0}=1.570796$ and its three harmonics. (b)
Non-stationary quasi-periodic (q-p) states for $\nu=0.30$ and
$T=0.35$ in terms of four basic frequencies.} \label{fig2}
\end{figure}

The spectrum in Fig. 2a clearly exhibits a period-4 solution. In
contrast, the spectrum of non-stationary solutions shown in Fig.
2b indicates quasi-periodic solutions. These are solutions
characterized by main frequencies that are linear combinations of
four recognizable basic frequencies.

\begin{figure}
\centering
\includegraphics[width=7cm,height=6cm]{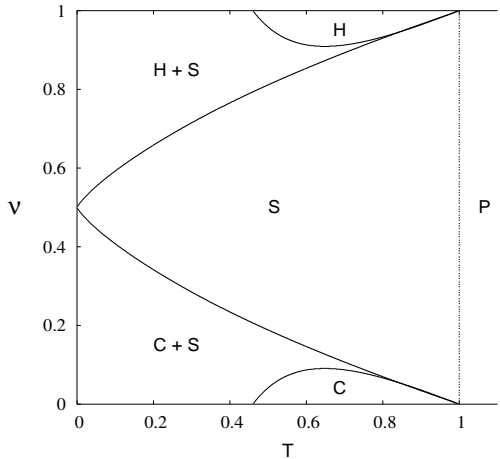}
\caption{Phase diagram for $c=3$ condensed patterns for $\alpha=0$
with initial overlaps $m_{1}=1,m_{2}=m_{3}=0$ and phases described
in the text.}
\label{fig3}
\end{figure}

Consider next the $(T,\nu)$ phase diagram of stable stationary
states for $\alpha=0$ and $c=3$, again with initial conditions
$m_1=1,m_2=m_3=0$ that favor single-pattern reconstruction, shown
in Fig. 3. The symmetric states are now stable over the whole
diagram, up to $T=1$, except for an upper and a symmetrically
placed lower region of unstable states, where Hopfield-like (H)
and cyclic (C) solutions, respectively as indicated, are the only
stable solutions. The stability of the symmetric solutions for all
$\nu$ at low $T$ is a characteristic feature of the phase diagrams
for an odd number of condensed patterns \cite{CS92}. There is also
a pair of tiny q-p solutions, that do not appear on the scale of
the figure, between the exclusive H and C regions and the
exclusive S phase for high $T$. The H or C solutions are stable
everywhere above or below the upper or the lower solid lines,
respectively. The symmetric states compete for stability with
Hopfield-like or cyclic solutions in the regions $H+S$ or $C+S$,
depending on initial values of the overlaps. The symmetric states
in the S phase are, again, the only stable states in the
intermediate region between the two first-order transitions. Other
initial more symmetric overlaps yield, again, smaller H and C
phases and pairs of q-p solutions between the S and the H and C
phases.

\subsection{Extensive loading}

For extensive loading, with $\alpha \neq 0$, one has to solve the
full set of Eqs. (11)-(17). We still expect to have three ordered
phases: H (Hopfield-like), S and C, with specific $\bm^*\neq 0$
and $q\neq 0$. Also, a disordered spin-glass phase (SG) should
appear with $\bm^*=0$ and $q\neq 0$ at the end of the ordered
phases and this phase should survive for all higher $T$. The
latter is a property of the layered network. All the transitions
are now expected to be discontinuous and all the phase boundaries
have to be obtained numerically.

One may expect the $\nu/(1-\nu)$ duality also to hold for non-zero
$\alpha$. Indeed, the Gaussian noise due to the non-condensed
patterns does not change neither under the set of linear
transformations that keep the probability distribution of the
condensed patterns invariant nor under the permutation index
matrix $\bS=\{S_{\mu \rho}\}$. These are the two basic
transformations which lead to a one-to-one correspondence between
every state in the upper part $(\nu>1/2)$ in the $(T,\nu)$ phase
diagram and a state in the lower part \cite{CS92}. The duality
serves to check the symmetry between the numerically constructed
phase boundaries.

The new feature of the $(T,\nu)$ phase diagrams for both $c=3$ and
$c=4$, not shown here, that are reached from Hopfield type initial
overlaps that favor single-pattern reconstruction,
$m_1=1,m_2=m_3=0$ and $m_1=1,m_2=m_3=m_4=0$, respectively, is that
the symmetric states become stable at low temperatures, including
$T=0$, within a finite range of intermediate values of $\nu$ as
soon as $\alpha$ is non-zero. These states remain as the only
stable states in that region, and there are H and C phases for
large and small values of $\nu$, respectively, and a SG phase at
high $T$. The phase boundaries to the H and C phases are nearly
the same for both $c=3$ and $c=4$, and there appear no q-p
solutions. The resulting phase diagrams are very similar for
either a Hebbian noise or full (Hebbian plus sequential) noise
with $b=\nu$, the main difference being that the symmetric phase
becomes somewhat enlarged at high $T$ in the latter case.

Other more symmetric initial overlaps, which favor symmetric
mixture states, lead to somewhat different results for $c=3$ and
$c=4$. For the latter, initial overlaps
$m_1=0.21,m_2=m_3=m_4=0.20$ reduce the size of the H and C regions
and lead now to intermediate regions of q-p solutions, as shown in
Fig. 4 for $\alpha=0.008$ and Hebbian noise. A very similar
diagram is obtained in the case of full noise with $b=\nu$, and
analysis of the power spectrum confirms the nature of the q-p
states as non-stationary quasi-periodic solutions. In contrast,
with initial overlaps $m_1=0.21,m_2=m_3=0.20$, in the case of
$c=3$, the symmetric states become the only stable states at low
$T$ for all values of $\nu$. Regions of stable Hopfield-like and
cyclic states appear at higher $T$ and, again, there appear
non-stationary q-p solutions between these phases and the $S$
phase.

\begin{figure}
\centering
\includegraphics[width=7cm,height=6cm]{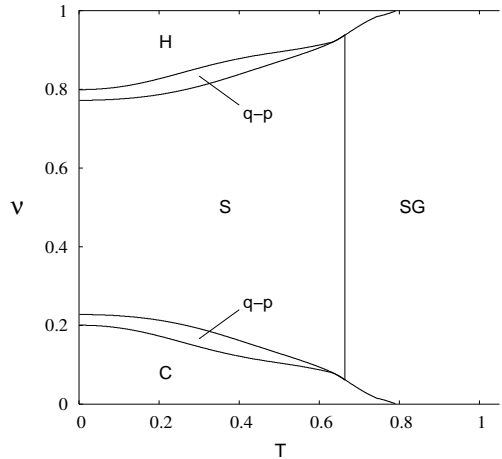}
\caption{Phase diagram for $c=4$, extensive loading
($\alpha=0.008$) and Hebbian noise, with initial overlaps
$m_{1}=0.21,m_{2}=m_{3}=m_{4}=0.20$. Spin-glass (SG) phase as
indicated.}
\label{fig4}
\end{figure}

All components of the overlap in the H phase are different from
zero for non-zero $\alpha$ at $T\neq 0$, although some of them may
be small. In contrast, some components may be zero when $T=0$, as
in the case of finite loading \cite{CS92}. Also, for non-zero
$\alpha$, the stable states in the C phase are still period-4
cyclic solutions. We checked this explicitly for $\alpha=0.008$
and values for $\nu$ and $T$ within that phase.

To see the overall role of extensive loading and to find the
critical storage ratio $\alpha_{c}$ for a given superposition of
Hebbian and sequential learning, we consider next the
$(\alpha,\nu)$ phase diagram of stable states at $T=0$ for $c=4$,
shown in Fig. 5, for Hebbian noise and initial overlaps
$m_1=1,m_2=m_3=m_4=0$ that favor single-pattern reconstruction.
Note that the critical storage ratio for pure Hebbian learning,
that is for $\nu=1$, is $\alpha_{c}=0.269$, in accordance with the
known result for the layered network model \cite{DKM89}. Due to
the $\nu/(1-\nu)$ duality, it is also seen to be the critical
ratio for pure sequential learning of the condensed patterns, as
it should be, since a pure sequential noise is a Hebbian noise.
The critical storage ratio $\alpha_{c}$ for the retrieval of
Hopfield-like or cyclic states is given by a point on the phase
boundaries where the H or C phases end. In order to check our
results obtained from numerical iterations, we performed numerical
simulations to locate a few points on the first-order transitions
between the ordered phases, at $T=0$ and we obtained results in
good agreement. As will be seen below (cf. Fig. 7), somewhat
different results are obtained in the case of full noise. Also,
more symmetric initial conditions, say
$m_1=0.21,m_2=m_3=m_4=0.20$, lead to further q-p solutions.

\begin{figure}
\centering
\includegraphics[width=7cm,height=6cm]{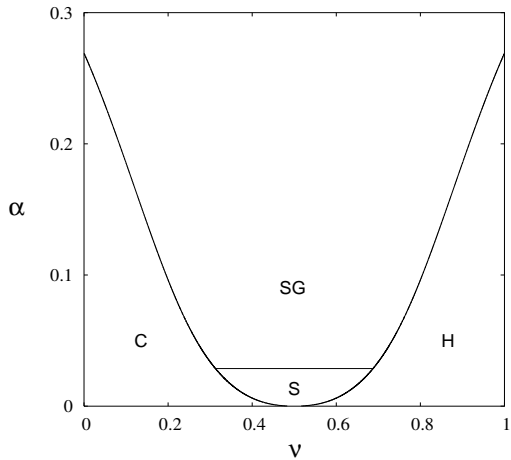}
\caption{Phase diagram for $c=4$, extensive loading at $T=0$ and
Hebbian noise, with initial overlaps
$m_{1}=1,m_{2}=m_{3}=m_{4}=0$.}
\label{fig5}
\end{figure}

\begin{figure}
\centering
\includegraphics[width=7cm,height=6cm]{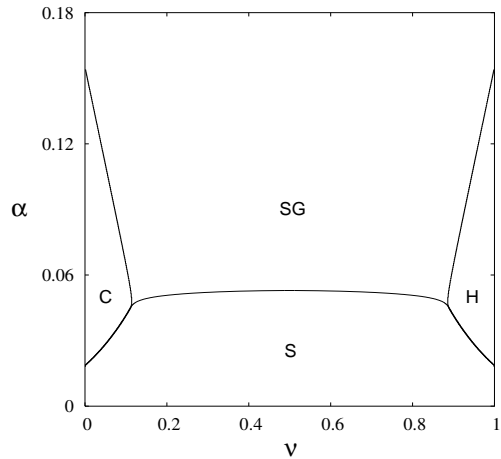}
\caption{Phase diagram for $c=3$, extensive loading at $T=0$ and
Hebbian noise, with initial overlaps
$m_{1}=0.4,m_{2}=0.3,m_{3}=0.2$.}
\label{fig6}
\end{figure}

We consider next the $(\alpha,\nu)$ phase diagram of stable states
at $T=0$ for $c=3$ and Hebbian noise. For the initial overlaps
$m_1=1,m_2=m_3=0$ one obtains a similar diagram to that shown in
Fig. 5 for $c=4$. The first-order transitions between the H and
the C phase with the SG phase are close to the case of $c=4$.
Instead, for more symmetric initial overlaps, say
$m_1=0.4,m_2=0.3,m_3=0.2$, one finds a quite different phase
diagram with a symmetric phase almost everywhere below a
transition to the SG phase and largely suppressed stable
Hopfield-like and cyclic states, as shown in Fig. 6. A similar
diagram, except for an enlarged S phase, is obtained in the case
of full noise with $b=\nu$.

We look now at the effects of Hebbian plus sequential noise for
$c=4$ and finite $\alpha$ at $T=0$. We do this again for $b=\nu$
and consider also the case of a matrix in which the high
components have equally favored Hebbian and sequential parts, in
order to explore the symmetry of the model around $b=0.5$,
independently of $\nu$. In each case we consider two kinds of
initial conditions: one favoring single pattern and cycle
retrieval, with $m_1=1,m_2=m_3=m_4=0$, and another more symmetric
with $m_1=0.21,m_2=m_3=m_4=0.20$.

\begin{figure}
\centering
\includegraphics[width=7cm,height=12cm]{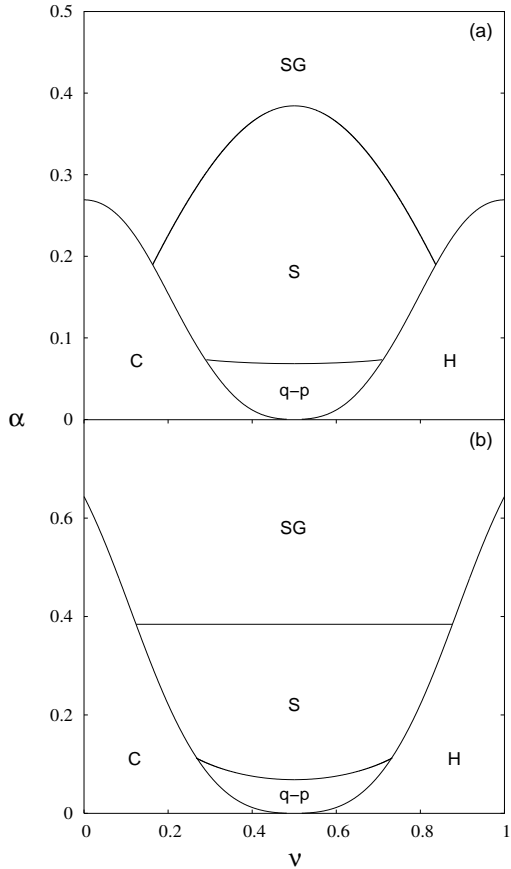}
\caption{Phase diagram for $c=4$, extensive loading at $T=0$, and
Hebbian plus sequential noise with initial overlaps
$m_{1}=1,m_{2}=m_{3}=m_{4}=0$, for (a) $b=\nu$ and (b) $b=0.5$.}
\label{fig7}
\end{figure}

The results for $b=\nu$ and $b=0.5$ for the first initial
condition are shown in Figs. 7(a) and (b), respectively. In both
cases there is now a finite region of non-stationary q-p solutions
below the phase of stable symmetric states, that was not present
for purely Hebbian noise. This is a feature of complete (Hebbian
plus sequential) noise and in the case of $b=0.5$ there is also a
considerable improvement of the critical $\alpha_c$ for
Hopfield-like and cyclic retrieval. Indeed, we find the quite
higher $\alpha_c=0.6438$ shown in Fig. 7(b). We also checked with
other choices for $b$ and found that $b=0.5$ seems to be the
optimal case.

Coming back to the more symmetric initial condition for $c=4$, one
finds quite larger regions of stable symmetric states and of
non-stationary q-p solutions and particularly smaller regions of
$H$ and $C$ phases for both $b=\nu$ and for $b=0.5$.

\section{Discussion and conclusions}

We studied in this work the dynamics of the competition between
pattern reconstruction and asymmetric sequence processing in an
exactly solvable feed-forward layered neural network model for
both finite and for extensive loading of patterns. The strictly
feed-forward nature of the network makes it an ideal system to
study the dynamics as a discrete-time evolution. Given the initial
overlaps as inputs on the sites of the first layer, the dynamics
is generated by random local fields at every site on the next
layer which depend on the synaptic interaction between units on
the two layers. In turn, the local fields determine the
probability of updating of units at the sites of that layer. The
asymmetry of the sequence processing in this work refers to the
sequential part of the learning rule that consists of a synaptic
interaction connecting a pattern on a layer with the consecutive
pattern in the sequence on the next layer. A symmetric sequence
processing, not considered in this work, would involve an
additional synaptic interaction of the same strength connecting a
pattern on a layer with the {\it previous} pattern in the sequence
on the next layer. We come back to this issue below.

The model studied here is based on the superposition of a Hebbian
and a sequential learning rule for a finite cycle of condensed
patterns and a stochastic noise due to a previously learnt
macroscopic set of either single patterns that follow a Hebbian
rule or a superposition of sequential patterns in a cycle with
single patterns. The superposition we consider is suitable for the
study of the competition between pattern reconstruction and
asymmetric sequence processing. We were especially interested in
the effects of finite synaptic noise ($\alpha\neq 0$) on the
performance phase diagrams that appear in the case of extensive
loading of patterns. Naively, one may expect that the form of the
noise does not make a great difference, as we found in most of
this work, but this is not always the case.

New dynamic equations for the overlaps and for the variance of the
Gaussian noise were obtained in the form of discrete-time
recursion relations for extensive loading of patterns. In the case
of a Hebbian plus sequential learning rule with a macroscopic
number of patterns $p=\alpha N$, the variance of the noise depends
on a macroscopic number of correlations between overlaps with
non-condensed patterns, $\langle\langle M^{\mu}_{N}(l)
M^{\rho}_{N}(l)\rangle\rangle$. All these correlations could be
relevant for the dynamics but when it comes to the fixed points
(and stationary cycles) only a small number of them survives
making the sequence processing a tractable and solvable problem,
at least in the relevant cases we considered in this work. In all
the cases we studied the correlations between overlaps with
non-condensed patterns formed, practically, a finite and tractable
set.

Explicit phase diagrams of stable states and regions of
non-stationary solutions were obtained in this work that show the
effects of stochastic noise due to a macroscopic number of learnt
patterns, and different behaviors may be obtained depending on a
variety of relevant parameters. We also considered, briefly, the
optimal case where $b=0.5$ in the synapses that generate the
noise. The reason why this yields a larger $\alpha_c$ for $\nu=1$
or zero is that the symmetric phase is considerably enhanced
towards extreme values of $\nu$ combined with the fact that
Hopfield-like retrieval in this model is, in general, more robust
to stochastic noise than the stable symmetric states for small or
large $\nu$.

Some general conclusions that may be drawn from the phase diagrams
are the following. First, as expected, the retrieval quality of
Hopfield-like and cyclic states is gradually reduced with an
increase in the storage ratio $\alpha$. Less obvious, there is
also a reduction in nearly single-pattern retrieval or in cycle
retrieval as $\nu$ or $1-\nu$ are decreased, respectively. The
change may be either to a symmetric state or to a spin-glass
state. Eventually, depending on the form of the noise due to the
previously learnt patterns, the transition could be to
non-stationary quasi-periodic states. The symmetric states appear
as a locking of the sequential transitions by the static Hebbian
part of the learning rule.

A further result of our work is that, for a Hopfield type initial
condition $m_1=1,m_2=m_3=m_4=0$, the first-order phase boundaries
in the ($T,\nu$) phase diagrams where the Hopfield-like phase ends
is practically the same for small $\alpha$ and either Hebbian or
complete noise, for both $c=4$ and for $c=3$. That is also the
case for the ($\alpha,\nu$) phase diagrams at $T=0$.

Turning now to predictions for somewhat larger, finite values of
$c$, either even or odd for Hopfield type initial conditions, we
expect that the non-stationary solutions that appear in the finite
loading case for small $T$ \cite{CS92} should disappear as soon as
$\alpha$ becomes non-zero, allowing for the presence of a
symmetric phase down to $T=0$, leaving a phase diagram with
essentially no q-p solutions.

It may be pointed out that there is a similarity between our
($\alpha,\nu$) phase diagrams at $T=0$, with initial Hopfield type
overlaps, and the corresponding phase diagram for pattern
reconstruction and {\it symmetric} sequence processing referred to
above obtained by means of statistical mechanics \cite{CT94}. The
case of pure sequential processing is not considered in that work
and, instead of a single-pattern retrieval region found there, we
have here a Hopfield-like region (a difference already noted in
the context of finite loading \cite{CS92}). The boundaries of
either of these regions have a similar dependence with the ratio
$\nu/(1-\nu)$. Moreover, there are similar regions of symmetric
phases and, in contrast to our work, there is also a region of
correlated states.

Based on the experience with the feed-forward layered network to
process information with a Hebbian learning rule, one may expect
qualitatively similar results to those obtained here to apply for
a recurrent network with a Hebbian-plus-sequential learning rule.
Most of the work reported here deals with the dynamics near
stationary states. Since our recursion relations for the overlaps
and for the noise are general for the layered network model, they
may also be employed to study the transients of the dynamics, in
particular to find out what makes the correlations between distant
(in pattern space) high components of the overlaps vanish. One may
also study the slowing down of the dynamics due to the macroscopic
number of non-condensed patterns. Work along some of these lines
is currently in progress.

As an extension of this work one may study the performance of this
model with a number of sequences. There is recent work only with
sequential learning, for an infinite number of limit cycles
\cite{MKO03}. We point out that there is renewed interest in
sequence learning from a biological point of view \cite{Ni03} and
that asymmetric networks have been found to be computationally
superior \cite{LD99}.\\

\section{Acknowledgments}

The work of one of the authors (WKT) was financially supported, in
part, by CNPq (Conselho Nacional de Desenvolvimento
Cient\'{\i}fico e Tecnol\'ogico), Brazil. Grants from CNPq and
FAPERGS (Funda\c{c}\~ao de Amparo \`a Pesquisa do Estado de Rio
Grande do Sul), Brazil, to the same author are gratefully
acknowledged. F. L. Metz acknowledges a fellowship from CNPq.

\end{document}